\documentclass{article}
\begin{document}
\markboth{G. F. Marranghello}
{Quark core formation in spinning-down pulsars}
\title{Quark core formation in spinning-down pulsars.}
\author{G. F. Marranghello, T. Regimbau, J. A. de Freitas Pacheco}
\maketitle
\begin{center}
Observatoire de la C\^ote d'Azur, Nice - France 
\end{center}

\begin{abstract}
Pulsars spin-down due to magnetic torque reducing its radius and increasing
the central energy density. Some pulsar which are born with central densities
close to the critical value of quark deconfinement may undergo a phase
transition and structural re-arrengement. This process may excite oscillation
modes and emmit gravitational waves. We determine the rate of quark core
formation in neutron stars using a realistic population synthesis code.
\end{abstract}

\section{Introduction}

Black holes ($BHs$) and neutron stars ($NSs$) are certainly two major
potential sources of gravitational waves ($GWs$). Unlike $BHs$, whose
gravitational waveforms are specified 
essentially by their masses and angular momenta, the characteristics of the
gravitational emission from $NSs$ depend on the properties of the nuclear
matter. 

Different mechanisms related to $NSs$ susceptible to produce large amounts of
$GWs$ have been investigated in the past years (see de Freitas Pacheco 2001
for a recent review). In particular, the mini gravitational collapse induced by
a phase transition in the core\cite{marranghello2}. If quark deconfinement
occurs in the central region of the $NS$, the core
will have a softer equation of state, inducing the system to search for a new 
equilibrium configuration, which will be more compact and having a larger
binding energy. The energy difference will partially cover the cost of the
phase transition and will be partially used to excite mechanical modes, which
will be damped either by heat dissipation or gravitational wave
emission. Radial 
oscillation modes are more likely to be excited after the mini-collapse if 
the star has a slow rotation. In this case, most of the mechanical energy 
will be dissipated in the form of heat and radiated away. If the $NS$ has an 
important rotation when the conditions for deconfinement are attained, then 
non-radial oscillations and radial modes coupled to rotation\cite{chau67} may
lead to an important $GW$ emission, whose energy amounts to about
$10^{52-53}\, erg$.

For a given equation of state for the hadronic and for the quark matter, the
deconfinement occurs when the Gibbs conditions are satisfied, e.g., equality
between pressure and chemical potential of both phases. If the baryonic mass 
of the configuration is high enough, the pressure and the energy density 
in the central regions attain values required for a phase transition to
occur. However, if the star is in rapid rotation, the central pressure
and energy density are below the critical values and the $NS$ has an internal
structure constituted essentially by hadrons. If, as expected, the $NS$ has
a magnetic field, the rotation velocity will decrease due to the canonical
magnetic dipole braking mechanism. Thus, after a certain time, the
phase transition conditions are reached and the star develops a quark core.

If one assumes that the above scenario is correct, then a natural question
appears. What is the expected frequency of these mini-collapse events ?
In a recent essay \cite{ma} based on very simple arguments and supposing
that  
Soft Gamma Repeaters or cosmological Gamma-Ray Bursts are a consequence of 
$NSs$ which have underwent a quark-hadron phase transition, estimated a frequency 
of $\sim 10^{-5}\,yr^{-1}$ per galaxy for these events. In the present work, a
more detailed estimate of the occurrence of these events is given. Rotating
$NS$ models have been computed for the equation of state derived by MVP02 in
order to estimated, for a given baryonic mass, what is the critical rotation
velocity below which the phase transition is possible. Then, using numerical
simulations as in \cite{regimbau}, the $NS$ flux (in
the $P$-$\dot P$ plane) crossing the critical region where the transition 
occurs was estimated and, as a consequence, the event frequency. 

\begin{center}
\begin{figure}[ht]
\vspace*{20pt}\vspace*{1.5truein}\parbox[h]{4.5cm}{\includegraphics{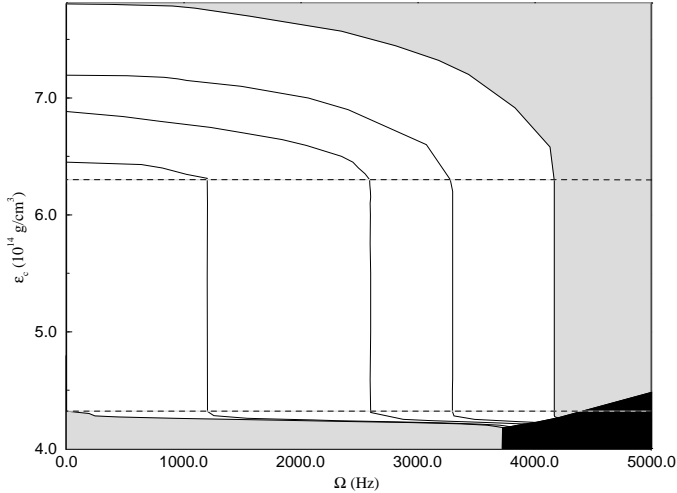}}\vspace{70pt}\caption{Evolutionary tracks for neutron stars
of different masses in the diagram $\epsilon_0 - \Omega$. Rotation periods
evolve according to the magnetic dipole braking mechanism.
In the upper shadowed region stars are already born with a quark core whereas in
the lower shadowed region, stars will never develop a quark core. The region
between the two horizontal lines corresponds to the energy density jump
during the phase transition. \label{f1}}
\end{figure}
\end{center}

Using the RNS rotation code\cite{stergioulas} we are able to compute the
evolution track of neutron stars with constant baryonic masses. Stars with
baryonic masses lower then 1.05 $M_\odot$, which represent 10$\%$ of the whole
population using a gaussian distribution centered in the 1.4 $M_\odot$, do
never form quark cores due to its low central pressure. In opposite, 70$\%$ of
the stars are already born with quark cores due o its high central
pressure. The figure \ref{f1} describes the stars with masses between these
two values and the formation of a quark core in a precise rotation frequency.

\section{The phase transition frequency}

As we have seen in the previous section, only $NSs$ born in the mass
interval $1.05 < M < 1.26$ will develop a deconfined core. For
a $NS$ of a given mass within that range, if its rotation velocity is not
zero, the central density will be below the critical value required
for the phase transition to occur. During its evolution the rotation
frequency decreases due to  magnetic torques and the central density
will eventually reaches the transition point. The timescale for
the occurrence of the transition depends on the initial rotation
period and magnetic field. The equation governing the number of
$NS$ in the rotation period $P$ space is
\begin{equation}
\frac{\partial N(P,t)}{\partial t} + \frac{\partial(\dot PN(P,t))}{\partial P} = S(P,t)
\end{equation}
where $N(P,t)$ is the number of $NSs$ at the instant $t$ with period $P$ in the
interval $P-(P+dP)$, $S(P,t)$ is the source function and $\dot P \equiv dP/dt
= f(P,H)$ describes the deceleration mechanism. Formally, in terms
of the Green's function, the solution of this equation can be written
\begin{equation}
N(P,t) = \int_{-\infty}^tdt_o\int dP_o~S(P_o,t)G(P-P_o,t-t_o)
\end{equation}
which depends on the initial distribution of periods ($P_o$) and
magnetic fields defining the evolution rate $dP/dt$. It worth mentioning
that, as consequence of this relation, the period $P$ is unambiguously
connected with the initial period $P_o$ at $t_o$.

In the present work, a different approach was adopted. We have
assumed that the initial distribution of rotation periods and
magnetic fields are independent on the NS mass, which obeys
a Gaussian distribution as we have already mentioned. Under these
conditions, the number of evolving $NSs$ that cross the interval
$P - (P+dP)$ is given by the current $J$ (Phinney \& Blandford 1981)

\begin {equation}
J=\frac{1}{\delta P}\sum\frac{dP}{dt}
\end{equation}

The current $J$ was calculated using the population synthesis code
developed by Regimbau \& de Freitas Pacheco(2001), up-graded to
take into account new pulsars discovered at high frequencies by the
Parkes Multibeam Survey and more recent models for the natal kick
distribution (\cite{arzoumanian01}). Pulsars are
generated at a constant rate using a Monte Carlo procedure and
their evolution in the Galaxy is followed during hundreds of million 
years. An interactive method permits modifications of the initial
distributions until observed distributions like those of the period,
period derivative, magnetic field, distances are well reproduced
when selection effects are taken into account (for details, the
reader is referred to \cite{regimbau}).
The derived parameters defining the properties of pulsars at birth are
given in table 4. The estimated number of ``active" pulsars in the
Galaxy is about 250,000, their birth rate is about one every 90 years
and their mean lifetime is about 22 million years.

\begin{center}
\begin{table}[ph]
\caption{Parameters of the initial period $P_0$ and magnetic braking timescale 
ln $\tau_0$ distribution, assumed to be Gaussian}
{\begin{tabular}{|cc|} 
\hline\hline
mean  &  dispersion  \\ 
\hline
$P_0 (ms) = 240 \pm 20$&$\sigma_{P_0} = 80 \pm 20$\\
$ln \tau_0 (s) = 11 \pm 0.5$&$\sigma_{ln\tau_0} = 3.6 \pm 0.2$\\ 
\hline
\end{tabular} \label{ta1}}
\end{table}
\end{center}

\begin{center}
\begin{figure}[ht]
\vspace*{20pt}\vspace*{1.5truein}\parbox[h]{4.5cm}{\includegraphics{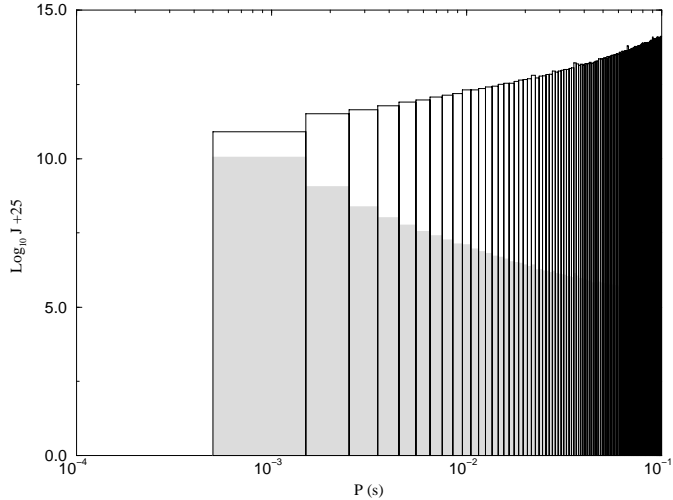}}\vspace{70pt}\caption{We show the neutron stars flow by second
    for each period and the flow of hybrid star formation in
    the filled bars.\label{fhy}} 
\end{figure}
\end{center}

In figure \ref{fhy} is shown the calculated current $J$ within bins of
$P-(P+\delta P)$. These bins are equivalent to mass bins $\delta M$ in
the interval $M_{min}-M_{sup}$ and the periods correspond to critical
values for which the transition occurs, which depends on the $NS$ mass.
The sum of the current in each bin, weighted by the number of pulsars
in that mass interval and averaged over the pulsar lifetime gives
a phase transition frequency of {\it $1.2\times 10^{-5}$} events per year.

\begin{center}
\begin{figure}[ht]
\vspace*{20pt}\vspace*{1.5truein}\parbox[h]{4.5cm}{\includegraphics{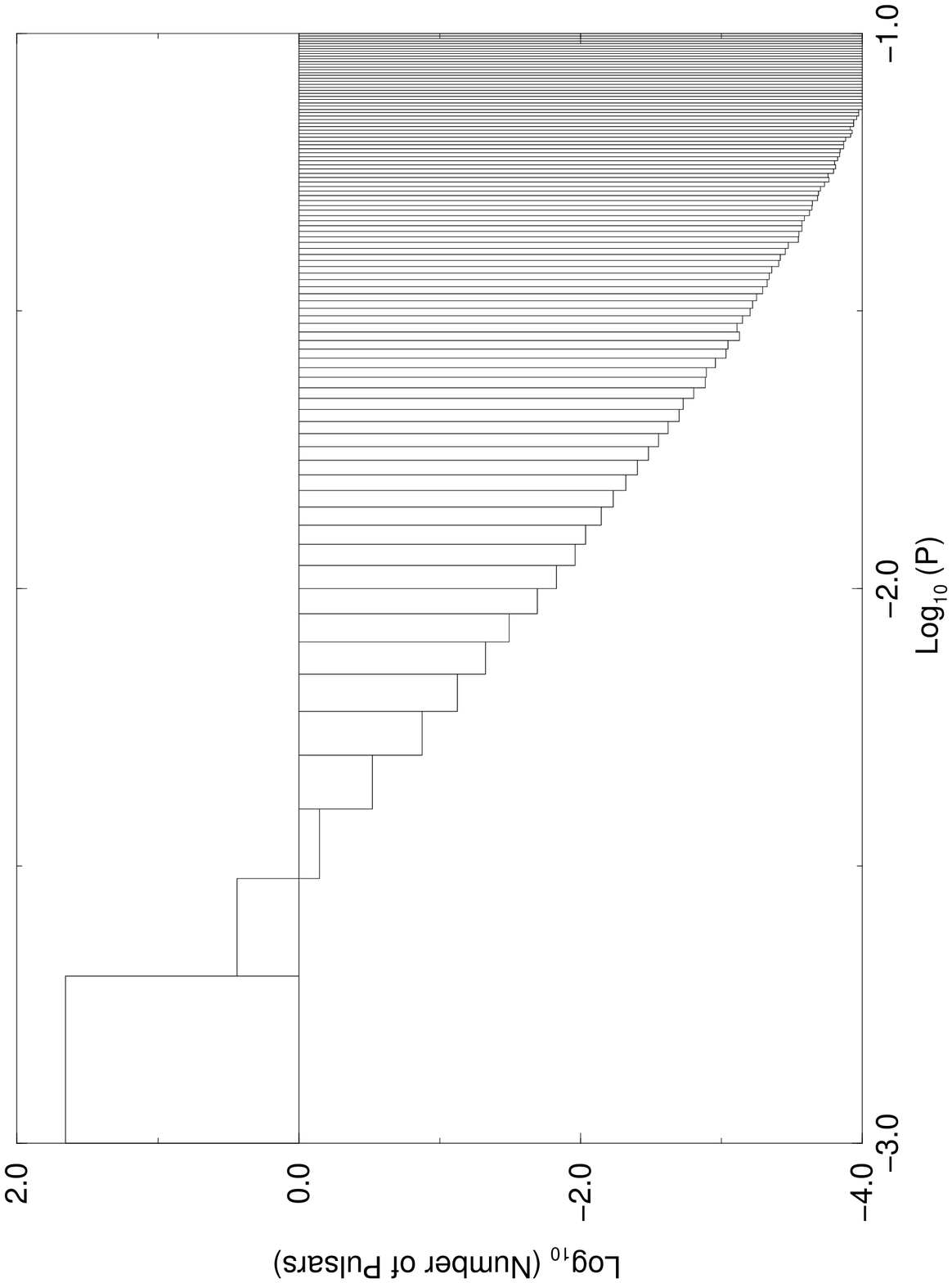}}\vspace{70pt}\caption{The number of pulsars which shall form a
    quark core as a function of period in our galaxy for a simulated
    population about 250.000 pulsars with an active life of 22$\times 10^6$yrs.\label{f2}}
\end{figure}
\end{center}

\section{Gravitational Waves}

Once we have determined the maximum distance probed (see \cite{marranghello2})
by gravitational wave 
detector and the rate of events in our galaxy, we assume that extragalactic
pulsars are formed with the same characteristics of those found in our
galaxy. We have extrapolated our results to extragalactic calculations using
the data found in (\cite{catalog,niklas95}) from where we extracted the
luminosity of close galaxy and 
clusters. We have used the luminosity as an indication of the formation of
stars and consequently of the formation of NS. The luminosity of the enclosed
members was divided by the Milky Way 
luminosity to derive the total number of detectable events. From
(\cite{catalog,niklas95}) we can extract, for example, the Local Group of
Galaxies, where our Galaxy is the most important, followed by M31 and M33. We
can also find at larger distances, objects like NGC672 (D$\sim$6 Mpc and
L$\sim 6.7\times10^{10}L_\odot$). Using this results we estimate the maximum
detection rate of only one event each 800 years for the present Virgo planned
sensitivity, which is low rate of events. This value get better by a factor of
2 if we are able to detect event up to 10 Mpc (Adv. Ligo planned sensitivity)
and reach one event each 100 years if the detectors are able to 
see this kind of events in the Virgo cluster.

\section{Conclusions}

Neutron stars evolution has been described. They were generated with periods
and masses that reproduces the characteristics of the detected pulsars. Than,
we have performed the spin-down due to magnetic torque, which increases the
central density of the stars. Some of these pulsars, borned with central
densities close to the deconfinement density, may undergo a phase transition
and suffer a micro-collapse. The rate such events were determined as well as
the possible rate of GW detection that come from them.

Most of the neutron stars cores are formed
during the supernovae event. The stars which form a quark core after the
supernovae explosion have their evolution determined by the initial mass and
period. Our model has predicted a rate of $10^{-5} events\cdot year^{-1}\cdot
galaxy^{-1}$. As these events can be detected by the gravitational wave
detectors for distances close to 7 Mpc (13 Mpc) by the Virgo (Adv. Ligo)
detectors, we can, in first approximation,
extrapolate the results obtained from
our galaxy to a greater distance which encloses more events and estimate a
rate of detection of about one each 800 years for the Virgo
detector. The planned advanced Ligo detector shall be able to see such events
close to the Virgo cluster and possibly detect one event each 100 years. This rate of events can still get better if one takes into account
the stars that undergo a phase transition due to mass accretion, however we do
not expect a great change in the actual value since there are only few
low-mass stars that could undergo the phase transition.

\end{document}